Spectrally Decomposed Denoising Diffusion Probabilistic Models for Generative Turbulence Super-Resolution

# Spectrally Decomposed Denoising Diffusion Probabilistic Models for Generative Turbulence Super-Resolution

M. Sardar,[1, a)] A. Skillen,[1] M. J. Zimoń,[2, 3] S. Draycott,[1] and A. Revell[1]
[1)]*School of Engineering,*
*The University of Manchester*
[2)]*IBM Research Europe*
[3)]*School of Mathematics,*
*The University of Manchester*

We investigate the statistical recovery of missing physics and turbulent phenomena in fluid flows using generative machine learning. Here we develop and test a two-stage super-resolution method using spectral filtering to restore the high-wavenumber components of two flows: a Kolmogorov flow and Rayleigh-Bénard convection. We include a rigorous examination of generated samples via systematic assessment of the statistical properties of turbulence. The present approach extends prior methods to augment an initial super-resolution with a conditional high-wavenumber generation stage. We demonstrate recovery of fields with statistically accurate turbulence on an $8\times$ upsampling task for both the Kolmogorov flow and Rayleigh-Bénard convection, significantly increasing the range of recovered wavenumbers from the initial super-resolution.

[a)]Electronic mail: mohammed.sardar@manchester.ac.uk

## I. INTRODUCTION

Numerical approaches to fluid dynamics typically require a compromise between computational accuracy and tractability. The broad range of length scales naturally occurring in turbulence generally means that lower-fidelity approaches necessarily lose some information – usually at the smallest scales. On the other hand, directly solving the governing equations remains computationally infeasible in most cases. Recent studies have demonstrated the potential to apply machine learning techniques to increase image resolution and recover missing details. So-called generative super-resolution methods are used to recover small-scale turbulent structures from fields produced by lower-fidelity methods. However, despite efforts to capture all scales through Super-Resolution (SR), the accurate reconstruction of small-scale turbulent structures remains elusive.

In low-resolution fields such as Large Eddy Simulation, the sub-grid scale stresses act to alter the resolved field, via the modelling of unresolved high-frequency turbulence information, which introduces errors and changes the trajectory of turbulence evolution. As a result, the solution field is not simply a low-pass filtered version of the high-fidelity field. Due to the non-linearity of turbulence, these small errors can become significant over time, which can make it difficult to account for lost information, or to infer useful information regarding the impact of unresolved turbulence on the resolved fields.

Fukami et al.[1] established that non-generative convolutional methods can be used to super-resolve turbulent flow fields by exploiting the ability of Convolutional Neural Networks (CNNs) to learn spatial correlations. Their analysis of vorticity distributions and spectra in the reconstructed flows revealed limitations to this approach; many super-resolved samples deviated substantially from the reference Direct Numerical Simulation (DNS). This was attributed to the fact that CNNs in SR produce blurred images without sharp detail. Attempts to address this were made by Liu et al.[2] by including a temporal component of the data in their super-resolution method. Taking contiguous-in-time snapshots of turbulence as training samples allows a CNN to include learnt temporal correlations. They found that this greatly improved the physical accuracy of their predictions, and reduced blurriness, but also reported poor performance in viscosity-dominated regions. The CNN method has also been successfully extended to 3D SR[3]. In more recent work regarding the CNN-based approach, Asaka et al.[4] have shown that combining wavelet projection techniques with CNNs and LSTMs can lead to enhanced CNN-based SR and time-series generation. However, when applied to Large-Eddy Simulation (LES) of shock waves, even recent CNN-based SR tends to underpredict the turbulent kinetic energy in areas of steep gradients[5].

The seminal work in super-resolution of turbulence using generative methods by Kim et al.[6] helped reinforce the idea that deep learning approaches can be powerful SR tools for turbulence. In particular they highlighted the success of generative methods in super-resolving data from low-resolution LES data, rather than low-resolution data obtained via downsampling of high-fidelity data. In doing so, they made us of a form of Generative Adversarial Network (GAN), and it was subsequently shown by Drygala et al.[7] that GANs can learn ergodic systems; i.e. wherein a single time evolution of the system will eventually traverse all possible states. The ergodicity hypothesis for turbulence is used frequently as justification for the equivalency between long-in-time time averaging and ensemble averaging approaches[8]. Drygala et al.[7] demonstrated a novel use of segmentation masks with a conditional GAN to synthesise LES samples of flow around a stator blade. This kind of conditioning information is useful in the context of Computational Fluid Dynamics (CFD) because it may be exploited to reproduce numerical boundary conditions and more accurately reproduce flows. Subsequent studies in this area have extended generative SR for turbulence: to more complex geometries[9], by introducing physical constraints[10,11], using alternative model architectures[12,13], and to spatiotemporal super-resolution[14].

Diffusion models are a novel form of generative method, exploiting useful properties of diffusion processes to generate samples of data[15]. They utilise learnt neural denoising functions to parameterise stochastic processes going from Gaussian noise to new samples of data. Diffusion models have the potential to provide superior quality generation, relative to GANs, while also avoiding several of the key challenges in training GANs, such as unstable training modes and mode collapse[16]. Diffusion models have demonstrated superior performance compared to GANs for image synthesis[17], albeit with an increased inference cost.

Inspired by recent developments in diffusion models using conditioning information for generation, Shu et al.[18] developed a physics-informed diffusion model for SR of turbulence. They demonstrated that incorporating physics residuals in the conditioning information given to the diffusion model during training can enhance the quality of generated samples as compared to a baseline SR Denoising Diffusion Probabilistic Model (DDPM)[15]. The work of Huang et al.[19] demonstrates that DDPMs without physics-conditioning are powerful SR models in 3D flows. Advances in DDPM-based SR demonstrate that this work can be extended to reduce inference cost and increase generative accuracy[20]. A key development in this area was made by Li et al.[21], wherein it was demonstrated that DDPMs outperformed GANs on an infill task[22], filling in patches of missing turbulence in 2D fields. Additional work by the authors includes DDPM-based generation of synthetic Lagrangian turbulence[23], and synthetic trajectories of Lagrangian turbulence[24]. DDPMs have also been applied to Eulerian turbulence time-series generation[25], and 3D flow field generation[26].

The present study aims to demonstrate a means of improving the statistical recovery of a turbulent flow field relative to the state of the art in the literature. Several derived quantities including the vorticity field, vorticity Probability Density Function (PDF), and the Turbulent Kinetic Energy (TKE) spectra are computed, to measure the statistical accuracy of the reproduced flow field. The spectral information is then used to define a cut-off for our second stage, high-wavenumber turbulence recovery. Initial turbulence recovery is accomplished by a spatial SR model[20], before a secondary recovery of the high-wavenumber structures is made, using the generated super-resolved low-wavenumber structures as conditioning information. To our knowledge, this is the first application of a 'continuous' DDPM to the super-resolution of turbulence. We anticipate that a two-stage approach can enable improved turbulence recovery to arbitrary wavenumbers, by making use of low-frequency information. We test this method on a Kolmogorov flow[27] and in a Rayleigh-Bénard Convection flow, both at an SR factor of $8\times$. We consider the $\mathrm{Ra} = 10^9$ case for the Rayleigh-B'enard Convection case due to its challenging nature and signficant fluctuations from the mean field. This is in contrast to prior SR studies on Rayleigh-Bénard Convection which have to-date considered lower Ra cases[28–30], where the turbulent fluctuations are less pronounced (with the notable exception being a recent study by Salim et al. [31], who considered a range of Ra up to $\mathrm{Ra} = 10^{10}$).

## II. METHODOLOGY

Here we detail our approach to turbulence recovery on two statistically stationary fluid flow problems: a Kolmogorov flow, and Rayleigh-Bénard convection. By integrating generative machine learning methods with a novel approach to deriving conditioning information from turbulent flow data, we aim to produce a two-stage generative model capable of generating statistically accurate snapshots of turbulence.

### A. Case 1: Kolmogorov Flow Data

Training and validation data are obtained from different snapshots of the same statistically stationary flow. The governing equations are the incompressible Navier-Stokes equations in 2D dimensionless form:

$$\begin{aligned} \frac{\partial u_i}{\partial t} + u_j \frac{\partial u_i}{\partial x_j} &= -\frac{\partial p}{\partial x_i} + \frac{1}{\mathrm{Re}} \frac{\partial^2 u_i}{\partial x_j \partial x_j} + f_i, \\ \frac{\partial u_i}{\partial x_i} &= 0, \end{aligned} \tag{1}$$

where $u_i$ is the dimensionless velocity field, $x_i$ is the spatial coordinate, $p$ is the dimensionless pressure, Re is the Reynolds number, $f_i = \sin(10\delta_{i2}x_1)$ is a steady sinusoidal forcing term, and $\delta$ is the Kronecker delta. The domain is taken as a square of length $2\pi$ in physical space, which for $256 \times 256$ grid points leads to a square of $128 \times 128$ in wavenumber space. The boundaries are fully periodic. We employ $Re = 222$ in our Kolmogorov Flow simulations, allowing for turbulent dynamics. The equations are solved numerically in the Fourier domain, using a spectral code[32]. The velocity fields are initialised uniformly based on a prescribed peak wavenumber, and are then filtered using white noise generated from a random seed to match a defined spectral density. Data from the initial stages of turbulence evolution is discarded so that the dataset contains only snapshots from the statistically stationary portion of the flow; this range is determined by computing the variation in time of the mean and variance of the velocity.

The autocorrelation function was used to define the decorrelation period when the autocorrelation function first passes through 0, allowing for the definition of a decorrelation write frequency. Using this, we generated 5000 samples of decorrelated training data. We reset the random seed for flow initialisation and then generated a further 600 samples of decorrelated validation data (also separated by the decorrelation period).

### B. Case 2: Rayleigh-Bénard Convection

Training and validation data are obtained from different snapshots of the statistically stationary portion of a Rayleigh-Bénard Convection (RBC) flow, discarding data during the initial RBC cell development. The governing equations are the incompressible Navier-Stokes equations in 2D, in non-dimensional form with the Boussinesq approximation for thermal-fluid coupling:

$$
\begin{aligned}
\frac{\partial u_i}{\partial t} + u_j \frac{\partial u_i}{\partial x_j} &= -\frac{\partial p}{\partial x_i} + \sqrt{\frac{\mathrm{Pr}}{\mathrm{Ra}}} \frac{\partial^2 u_i}{\partial x_j \partial x_j} + T^* \delta_{i2} \hat{x}_2, \\
\frac{\partial T^*}{\partial t} + u_i \frac{\partial T^*}{\partial x_i} &= \sqrt{\frac{1}{\mathrm{RaPr}}} \frac{\partial^2 T^*}{\partial x_i x_i}, \\
\frac{\partial u_i}{\partial x_i} &= 0,
\end{aligned} \tag{2}
$$

where $u_i$ is the velocity field, $x_i$ is the spatial coordinate, $p$ is the pressure, $T^*$ is a scalar temperature field, Pr is the Prandtl number, Ra is the Rayleigh number, $\hat{x}_2$ is a unit vector in the vertical direction, and $\delta$ is the Kronecker delta. The Rayleigh number is defined as $Ra = \frac{\beta \Delta T l^3 g}{\nu \alpha}$, where $\beta$ is the thermal expansion coefficient, $\Delta T$ is the temperature difference between the hot and cold walls, which are separated by $l$, $g$ takes its usual meaning as gravitational acceleration, $\nu$ is the kinematic viscosity, and $\alpha$ is the thermal diffusivity. The Prandtl number is defined as $Pr = \frac{\nu}{\alpha}$, where $\nu$ is the kinematic viscosity, and $\alpha$ is the thermal diffusivity as before. The domain is taken as a rectangle of aspect ratio 2, with unit height, discretized into $512 \times 1024$ spatial grid points, resolving to approximately $3l_k$, where $l_k$ is the Kolmogorov length scale. Streamwise boundaries are periodic, and the upper and lower boundaries have prescribed temperatures. We employ $\mathrm{Pr} = 1$ and $\mathrm{Ra} = 10^9$ in our simulations, allowing for turbulent dynamics. We note that this is higher than other SR analyses of the RBC problem, and leads to complex physics. The equations are discretized into Fourier space in the streamwise (periodic) direction, and Chebyshev bases are used in non-periodic directions, using a spectral code[33]. Pressure fields are initialised by generating a field using filtered Gaussian noise on a grid, from which the initial velocity and buoyancy fields are computed.

Using the decorrelation write frequency, we generated 2400 samples of decorrelated training data using 8 random seeds. We reset the random seed for flow initialisation and then generated a further 300 samples of decorrelated validation data (also separated by the decorrelation period). We do not downsample the high-resolution data (a technique often used to make training more tractable[34,35]). Downsampling leads to truncation of the wavenumbers represented in the data, effectively cutting off the dissipation range of the power spectral density.

### C. Diffusion Model

A DDPM relies on learning the reverse process induced by a Markov chain of Gaussian transitions which act to iteratively add noise to data, transforming a distribution of data to a Gaussian. By learning the reverse process, a diffusion model can generate samples of data from pure noise. In discrete-time, the forward diffusion process of corrupting data is completed via a 'variance schedule', which defines a piece-wise smooth transition from the data to Gaussian noise. The continuous-time DDPM formulation presented in Ho and Salimans [20] allows for a continuous parameterization of the forward process by conditioning directly on $\lambda$, the log Signal-to-Noise Ratio (SNR). This is given one of several functional forms, corresponding to various noising schedule strategies. The simplest of these is a linear relationship between the limits of $\{\lambda_{min}, \lambda_{max}\}$. Ho and Salimans [20] demonstrate that the noising process is entirely parameterised by these two limits. Figure 1 presents two choices of $\{\lambda_{min}, \lambda_{max}\}$ alongside samples noised to different $\lambda^*$ between them, one in which information is destroyed quickly, and one in which the information is destroyed more gradually. Here, $\lambda^*$ is used to denote $\lambda$ normalised between 0 and 1. Note that the choice of $\{\lambda_{min}, \lambda_{max}\}$ is not required to be symmetric around 0, and is a hyperparameter.

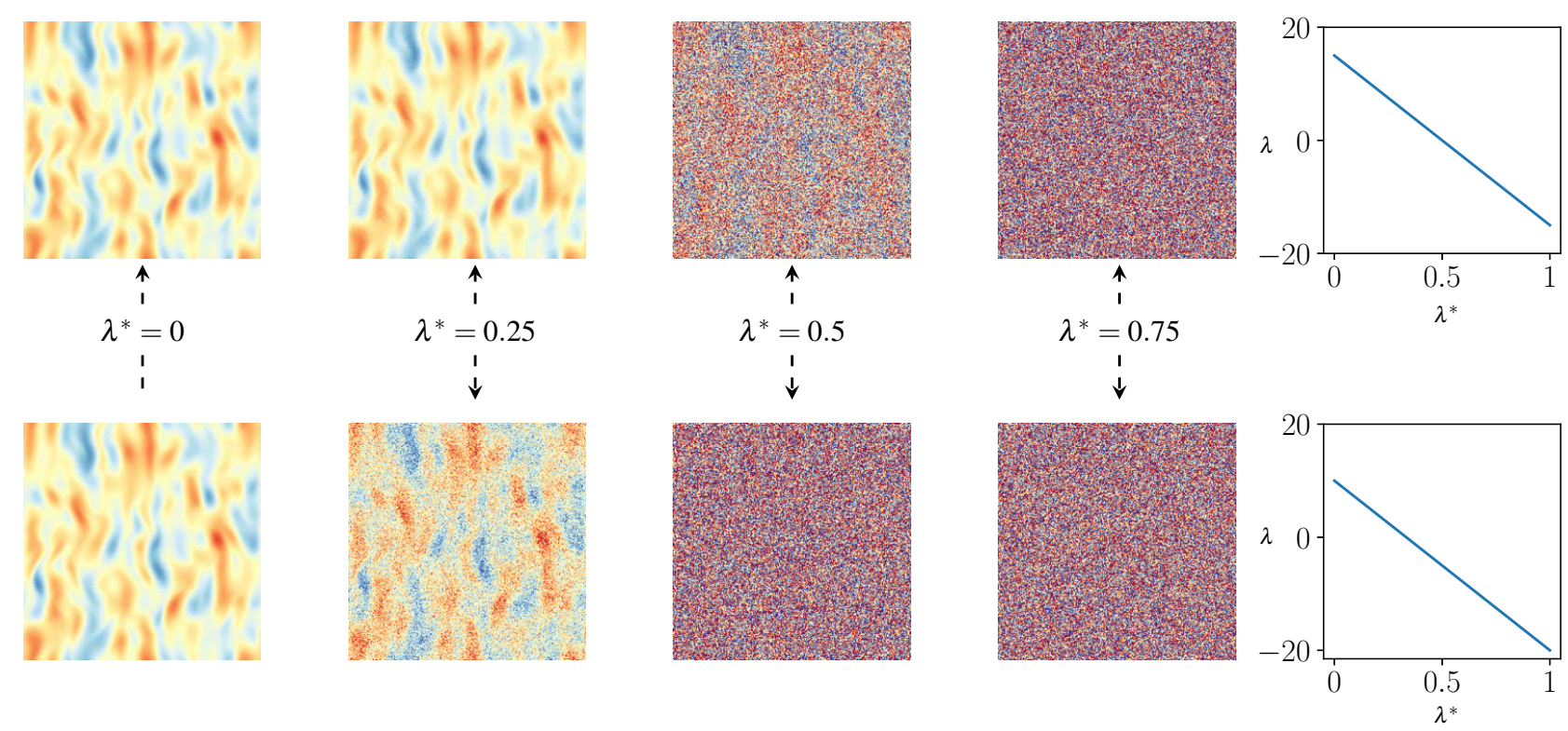

**FIG. 1:** A comparison of two choices of $\{\lambda_{min}, \lambda_{max}\}$ in terms of their impact on a sample from the Kolmogorov flow dataset.

We use a DDPM[15] with classifier-free-guidance[20] to super-resolve an average-pooled velocity field for Kolmogorov flow, and an average-pooled temperature field for the RBC flow. A UNet[36] CNN is used as the backbone of our diffusion model. During training, we show the diffusion model Low-Resolution (LR) average-pooled DNS samples and corresponding noisy High-Resolution (HR) DNS samples from a given instance of DNS for spatial super-resolution. High-resolution samples are put through the forward noising process (Figure 1), and concatenated to their counterpart low-resolution samples. This is fed through the UNet, which is trained to predict the noise added to the HR sample while also suitably conditioning on the low resolution information available. In effect, the UNet is trained to be a conditional denoising function. From[20], we additionally discard the conditioning information with a probability of $p_{uncond}$, which we specify as 0.2. This is used for the classifier-free guidance, which is discussed below.

At inference time, the UNet is used as one component of an iterative scheme for generation. Generation is carried out by discretizing the functional form of the log(SNR) using the limits of $\{\lambda_{min}, \lambda_{max}\}$ chosen during training, and iteratively traverses the associated reverse Markov chain to generate a new sample from the original data distribution, conditional on the low-resolution sample (i.e. super-resolution). The reverse process is defined as $q(z_{\lambda'}|z_\lambda, x) = \mathcal{N}\left(\tilde{\mu}_{\lambda'|\lambda}(z_\lambda, x), \tilde{\sigma}^2_{\lambda'|\lambda} I\right)$, with terms defined in Eq. 3[20]:

$$\tilde{\mu}_{\lambda'|\lambda} = e^{\lambda-\lambda'}\left(\frac{\alpha_{\lambda'}}{\alpha_\lambda}\right) z_\lambda + (1-e^{\lambda-\lambda'})\alpha_{\lambda'} x, \quad \tilde{\sigma}^2_{\lambda'|\lambda} = (1-e^{\lambda-\lambda'})\sigma^2_{\lambda'}, \tag{3}$$

where $\tilde{\mu}_{\lambda'|\lambda}$ is the reverse process mean at some $\lambda' < \lambda$ given the mean at $\lambda$, $\alpha_\lambda = \sqrt{1/\left(1+e^{-\lambda}\right)}$, $z_\lambda$ represents a sample from the target distribution noised to $\lambda$, $x$ represents the original noise-free high-resolution image, $\sigma^2_\lambda$ is the forward process variance at $\lambda$, and $\tilde{\sigma}^2_{\lambda'|\lambda}$ is the reverse process variance at some $\lambda' < \lambda$.

As our process is generative, $x$ is instead parameterised by the trained UNet, such that $x_\theta(z_\lambda) = (z_\lambda - \sigma_\lambda \varepsilon_\theta(z_\lambda, c, \lambda))/\alpha_\lambda$, where $\varepsilon_\theta(z_\lambda, c, \lambda)$ is our trained UNet, given an input noisy image ($z_\lambda$), conditioning information ($c$) in the form of a low-resolution image, and $\lambda$. The conditioning tensor is not required to be a low-resolution image, a fact we exploit in our proposed 'second-stage generation', to be discussed.

As mentioned above, the UNet is trained partially on the unconditional generation task. This is then used during inference, with a guidance strength parameter, $w$, to obtain an interpolated value for the noise prediction during inference, $\tilde{\varepsilon}_\theta = (1+w)\varepsilon_\theta(z_\lambda, c, \lambda) - w\varepsilon_\theta(z_\lambda, \lambda)$.

The initial generative process starts at $p_\theta\left(z_{\lambda_{min}}\right) = \mathcal{N}(0, I)$. The transitions defined in Eq. 3 then become Eq. 4:

$$p_\theta(z_{\lambda'}|z_\lambda) = \mathcal{N}\left(\tilde{\mu}_{\lambda'|\lambda}(z_\lambda, x_\theta), \left(\tilde{\sigma}^2_{\lambda'|\lambda}\right)^{1-v}\left(\sigma^2_{\lambda|\lambda'}\right)^{v}\right), \tag{4}$$

where, for classifier-free guidance during inference, $v$ gives an interpolation strength between the variance for unconditional and conditional generation. Hyperparameters $w$ and $v$ were set for each case based on a random search. Since these are only used during inference, the cost of this search was relatively low. This hyperparameter search is case-dependent; we note that for the Kolmogorov flow, lower values of $w$ were favourable, while the RBC case favoured higher values of $w$ for accurate reconstruction of turbulence statistics.

### D. Fourier Filtering

Fourier transforms allow for analysis of turbulence in wavenumber space. We consider only discrete Fourier transform, computed using Fast Fourier Transform (FFT). The maximum wavenumber is the Nyquist wavenumber, defined as $k_{max} = \pi N/L$ for N uniformly spaced grid points along a given axis. Here, as $L = 2\pi$, $k_{max} = 1/2N$. The discrete 2D FFT, applied to velocity, is given in Eq. 5, and the wavenumber is the magnitude of the wavevector is given in Eq. 6:

$$\mathscr{F}\left(\mathscr{F}\left(u_i\right)_x\right)_y = \sum_{m=0}^{N_y-1}\sum_{j=0}^{N_x-1} u_i e^{-2\pi \mathrm{i}\left(j/N_x + m/N_y\right)}, \tag{5}$$

$$k = |\boldsymbol{\kappa}| = \left|\begin{matrix}\kappa_x \\ \kappa_y\end{matrix}\right| \quad \begin{matrix}\text{for } -\frac{1}{2}N_x \leq \kappa_x \leq \frac{1}{2}N_x, \\ \text{for } -\frac{1}{2}N_y \leq \kappa_y \leq \frac{1}{2}N_y,\end{matrix} \tag{6}$$

where $N_i$ is the number of grid points in each spatial direction, and $u_i$ is used to denote separate treatment of the longitudinal and transverse velocity components. We select a circular filter in wavenumber space with a radius equal to the wavenumber limit that is resolved by the first-pass SR.

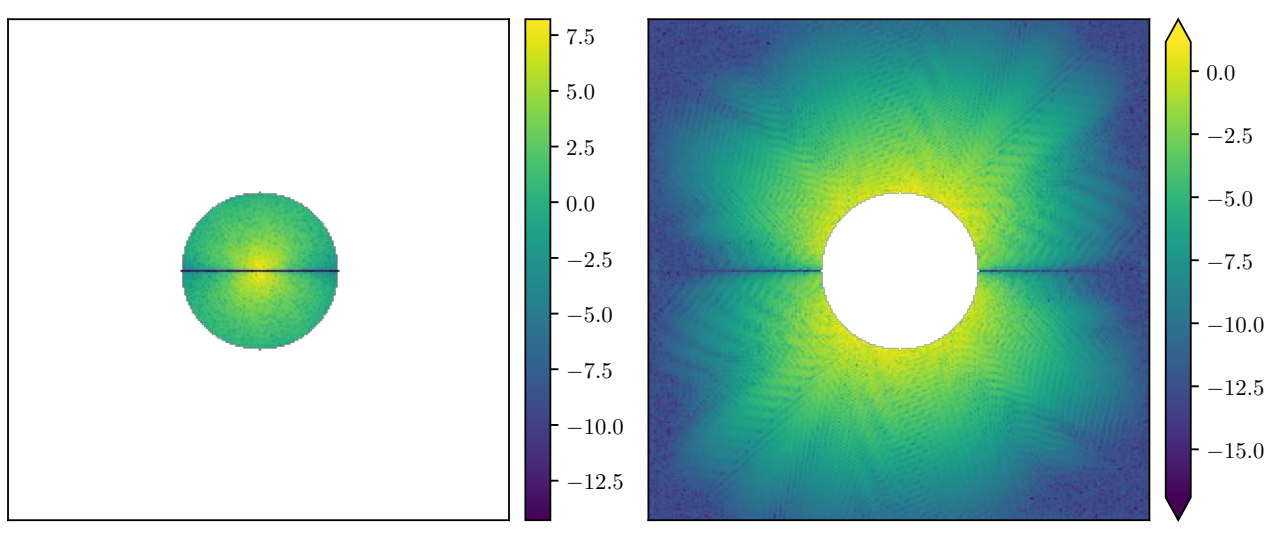

**FIG. 2:** Spectral filters applied to a given 2D sample of the Kolmogorov flow, at $k \leq 40$ and $k > 40$. The field shown here is $\log\left(\left|\mathscr{F}\left(\mathscr{F}\left(u_x\right)_x\right)_y\right|\right)$, which is the log magnitude of the 2D FFT amplitude.

Figure 2 shows a given sample of turbulent flow passed through a low-pass filter and high-pass filter at $k = 40$ in wavenumber space. These samples are then transformed back to physical space using a 2D inverse FFT (Figure 3). We include additional filtering to highlight the relative orders of magnitude between wavenumber ranges. We hypothesise that this may contribute to difficulties encountered in capturing higher-frequency components in super-resolved turbulence[1,6,37]; p-norm based loss functions in ML approach converge on orders of magnitude significantly lower than the amplitude of these wavenumbers in physical space, which makes recovery of information beyond this difficult.

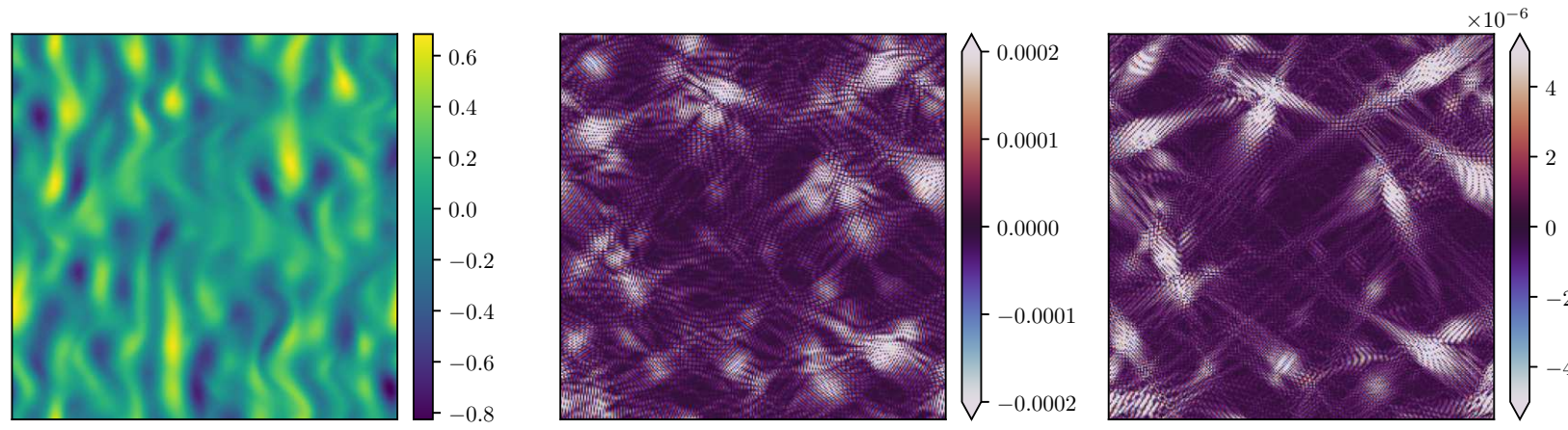

**FIG. 3:** $u_x$, passed through spectral filters: $k \leq 64$ (left), $k > 64$ (center), $k > 96$ (right). We note the changing magnitude of each field, illustrating the numerical difficulty in capturing these components. Colour bars for the High-Pass Filtered (HPF) samples are scaled to better illustrate dominant fluctuations.

Fourier-filtering provides a new dataset used for training the stage two model, i.e., the spectral-filtered diffusion model. Once the first-stage super-resolution model is trained and evaluated, we analyse the Turbulent Kinetic Energy (TKE) spectra of generated SR samples to heuristically determine the wavenumber after which the SR spectrum begins to deviate from the DNS. This wavenumber is denoted as $k_{filter}$, and is used to filter the DNS into low-wavenumber and high-wavenumber components. During training, the model is shown the velocity fields passed through a Low-Pass Filter (LPF), and High-Pass Filter (HPF), and trained to predict noise content in high-frequency fields conditioned on low-frequency information. In a similar approach to that outlined above, given pure Gaussian noise and a low-frequency field obtained by filtering the first stage super-resolution, the model aims to generate new samples of high-frequency complement information conditional on the generated low-frequency fields. The generated high-wavenumber fields are then analysed. Lastly, the generated high-wavenumber fields are superimposed onto the fields output from the stage 1 super-resolution, by first filtering out the incorrect information in the stage 1 generated fields using the same spectral filtering, then adding the high-wavenumber generated fields corresponding to the same ground-truth DNS sample.

The full dataset preparation, training, and inference process is detailed in Figure 4.

## III. RESULTS

### A. Case 1: Spatial Super-Resolution of a Kolmogorov Flow

Here we present the results of a spatial super-resolution diffusion model trained to learn the distribution of high-resolution fields, $u_i$, given low-resolution representations of the fields. We train a DDPM (specifically a continuous-time DDPM[20]) to reconstruct a $256 \times 256$ field from a $32 \times 32$ low-resolution sample obtained by average-pooling the DNS solution. We examine a variety of quantitative and qualitative metrics to determine the extent to which turbulence has been recovered in our approach, on a validation dataset.

The vorticity, $\omega = \nabla \times \mathbf{u}$, is computed using an 8th-order central difference in physical space for the high-resolution and a 2nd-order approximation for the low-resolution. We should anticipate an approximately symmetric distribution of vorticity centred at 0 for individual samples.

From Figure 5, we observe that after the first stage of super-resolution, on an $8\times$ upsampling task from $32 \times 32$ to $256 \times 256$ points, the flow structures are recovered well. The characteristic interactions between positive and negative vortices are also captured well, with clockwise and counterclockwise rotating structures. We note that our aim in generative super-resolution is not to produce a like-for-like version of the high-resolution sample, but to learn an underlying statistical description of the high-resolution data. To this end, some perceptual difference is acceptable, provided the generated and ground truth samples are statistically similar.

The treatment of flow variables as continuous random variables implies that their PDF completely characterises them – rather than deterministically attempting to predict the value at a timestep, it is sensible to consider the likelihood of a variable taking a certain value. We show the PDF of an instantaenous vorticity snapshot in Figure 6. We observe that the super-resolved data shows good agreement with the high-resolution data. This demonstrates the super-resolved fields are statistically similar to the ground truth DNS, and present a significant improvement in the primary statistics from the LR data.

In order to investigate the recovery of information at all length scales of turbulence, we must examine the turbulent kinetic energy spectra. Figure 7 shows that through a learnt $8\times$ upsampling, the spatial SR diffusion model is able to recover wavenumbers in the range $16 < k \leq 40$, after which the SR solution diverges. Typically in CNN-based SR approaches, we expect an underprediction of the spectrum[5,37], beyond certain wavenumbers. Results from[18] indicate diffusion-based recovery to higher wavenumbers than $k = 40$, which motivates the use of their physics-guided DDPM method. However, we note that the level of correlation in their dataset was not discussed, which may affect the validity of their test data. The recovered wavenumber limit is shown with a purple dashed line. The missing information beyond $k = 40$ may be explained by the numerical limitations in learning information at the high wavenumbers (§II D), and limited model capacity (due to training limitations). In a DDPM, missing information beyond a certain wavenumber is observed via flatlining of the spectra due to residual noise beyond that scale, as per Dieleman[38].

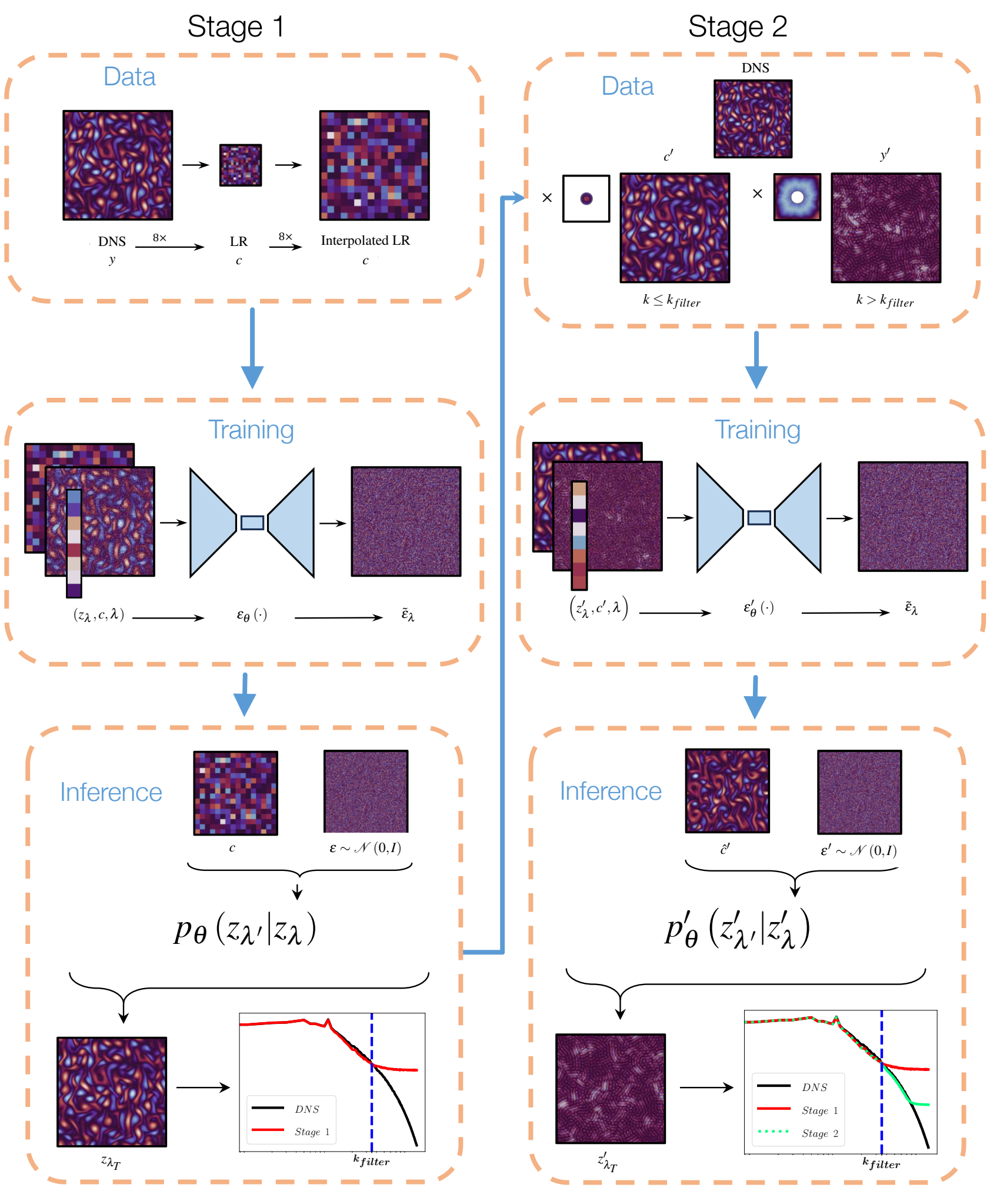

**FIG. 4:** Procedure for spectrally decomposed turbulence generation. In stage 1, DNS data is downsampled to produce pairs of low and high resolution samples $(c, y)$. A DDPM, $\varepsilon_\theta$ is trained to generate upsampled snapshots of turbulent flow. TKE spectra of the generated samples are compared against those of the DNS, giving a 'cutoff' wavenumber, $k_{filter}$. In stage 2, $k_{filter}$ is used to filter the DNS in wavenumber space to obtain pairs of low and high pass filtered velocity fields, $(c', y')$. These are then used to train a second DDPM, $\varepsilon'_\theta$. This may then be used in inference to obtain a generated high-pass filtered field, conditional on the low-pass filtered field from stage 1 generation.

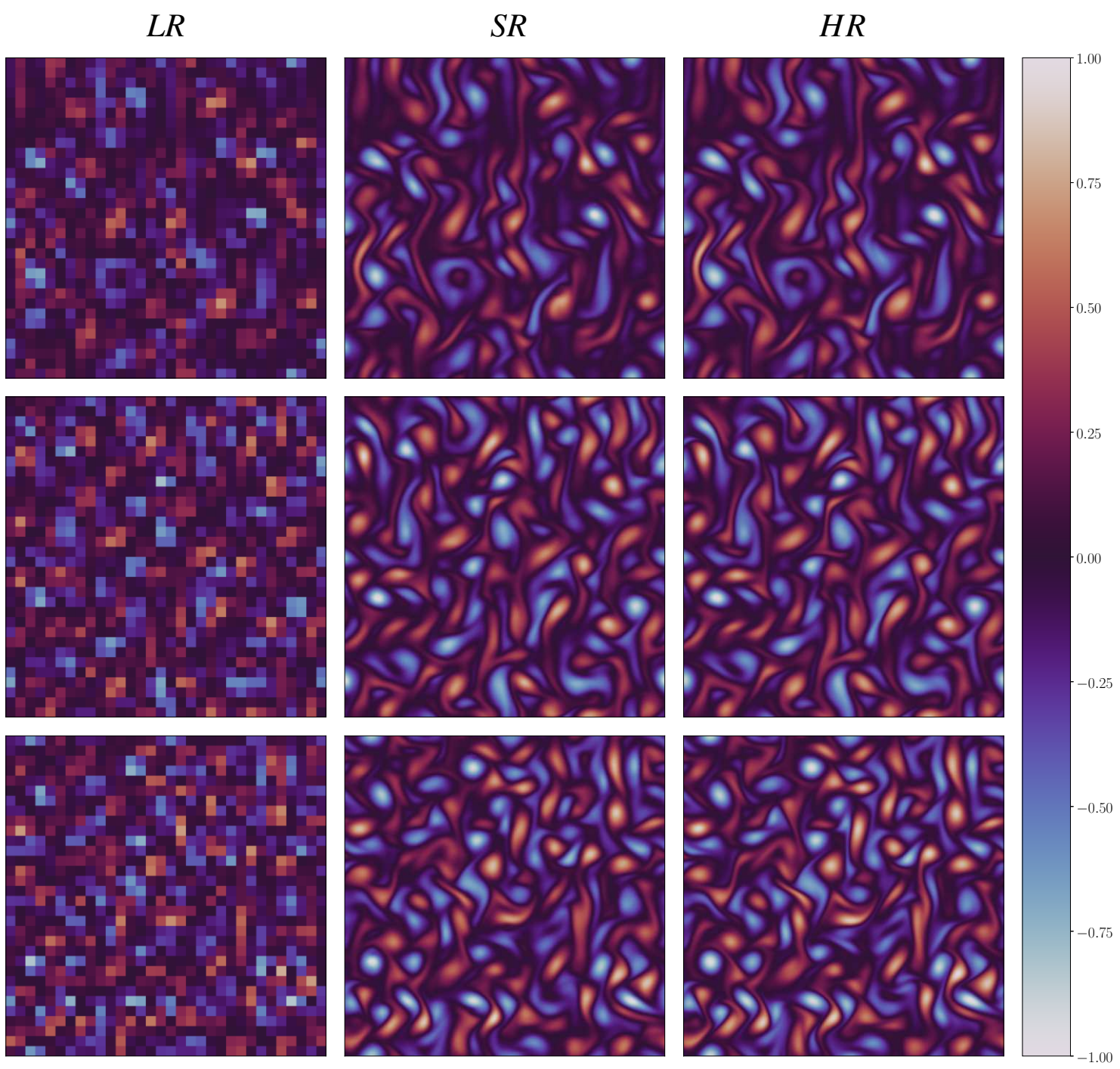

**FIG. 5:** Contours of $\omega^* = \dfrac{\omega}{\omega_{max}}$ for the first stage spatial super-resolution, for three representative samples. The $32 \times 32$ [left] is super-resolved using a diffusion model to $256 \times 256$ [middle], which we compare to the ground truth DNS field [right].

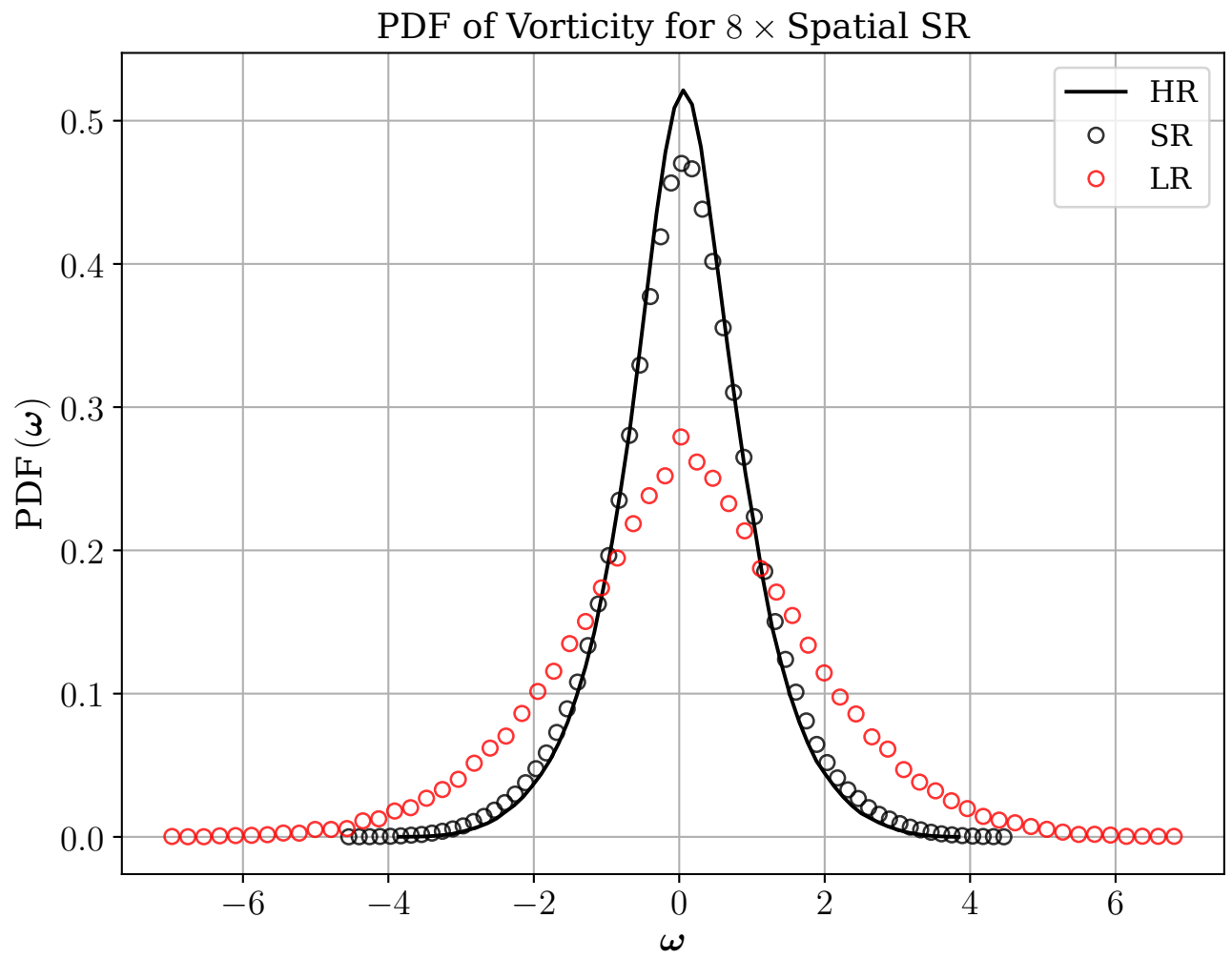

**FIG. 6:** PDF$(\omega)$ for the first stage, spatial super-resolution. LR is $32 \times 32$, HR is the $256 \times 256$ DNS, and SR is from the super-resolved fields.

### B. Case 1: Fourier-Filtered Turbulence Generation

It is evident that the high wavenumber turbulent structures are difficult to capture due to their small contribution to the flow structure, notwithstanding their significant contribution to the flow statistics. Both perceptually and from the probability distribution, it seems that turbulence recovery up to $k = 40$ appears to be sufficient for reasonable first order statistics in this case. However, spectral analysis has shown that there is still non-negligible energy within these high wavenumbers, and for the purposes of scale-resolving simulations, these wavenumbers are considered significant. Here we demonstrate that using information from the spectrum of the spatially super-resolved flow, a second stage of turbulence recovery can be executed using a diffusion model trained to generate high-wavenumber flow components conditioned on their low-wavenumber complement.

From Figure 7, we identify the line at $k = 40$ as the wavenumber at which the super-resolved field starts to diverge from the reference DNS solution. The recovery up to this wavenumber is strong. As such, we take a spectral filter (Figure 2) and apply it to the 2D FFT of the velocity field, i.e., the velocity field in wavenumber space. We then convert the fields back to physical space, obtaining low-wavenumber, high-wavenumber pairs as in Figure 8. The conversion back to physical space is required as the generative model is not guaranteed to produce real (i.e. non-complex) fields if generating in wavenumber space. A key feature of this form of filtering operation is that the sum of the low-wavenumber and high-wavenumber field in physical space is equal to the ground truth DNS field; this is exploited in order to use this approach for high-wavenumber correction of the spatial super-resolution.

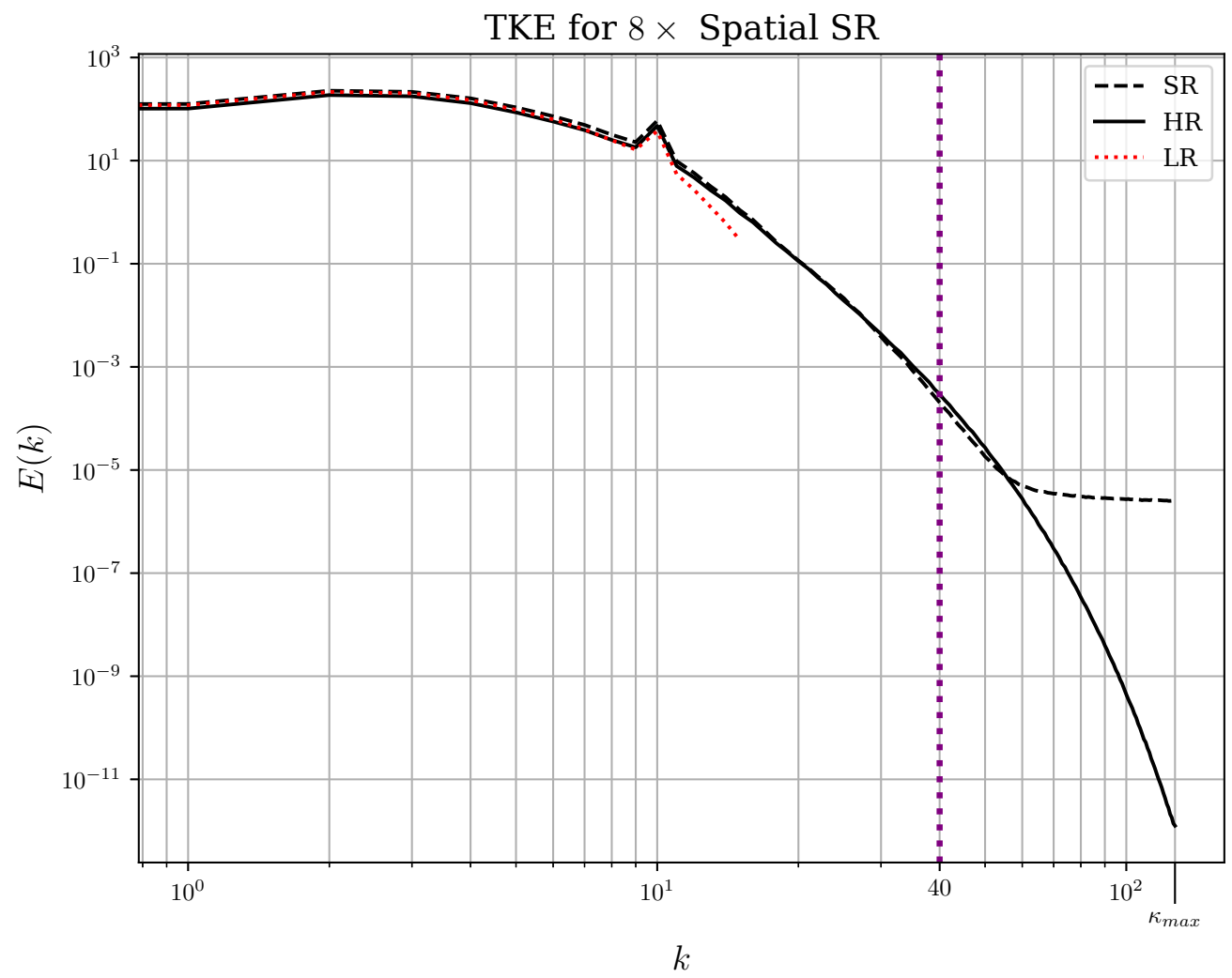

**FIG. 7:** TKE Spectra from the single-stage super-resolution. *LR* is $32\times32$, *HR* is the $256\times256$ DNS, and *SR* is from the super-resolved fields.

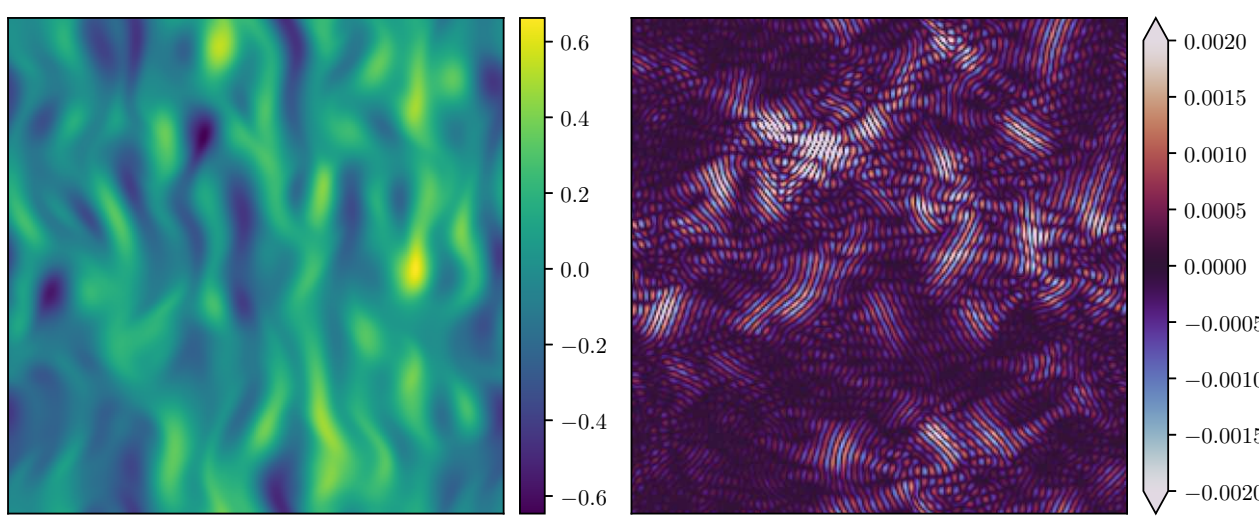

**FIG. 8:** Representative sample of $u_x$ with $k \leq 40$ (left), and $k > 40$ (right), prior to standardisation.

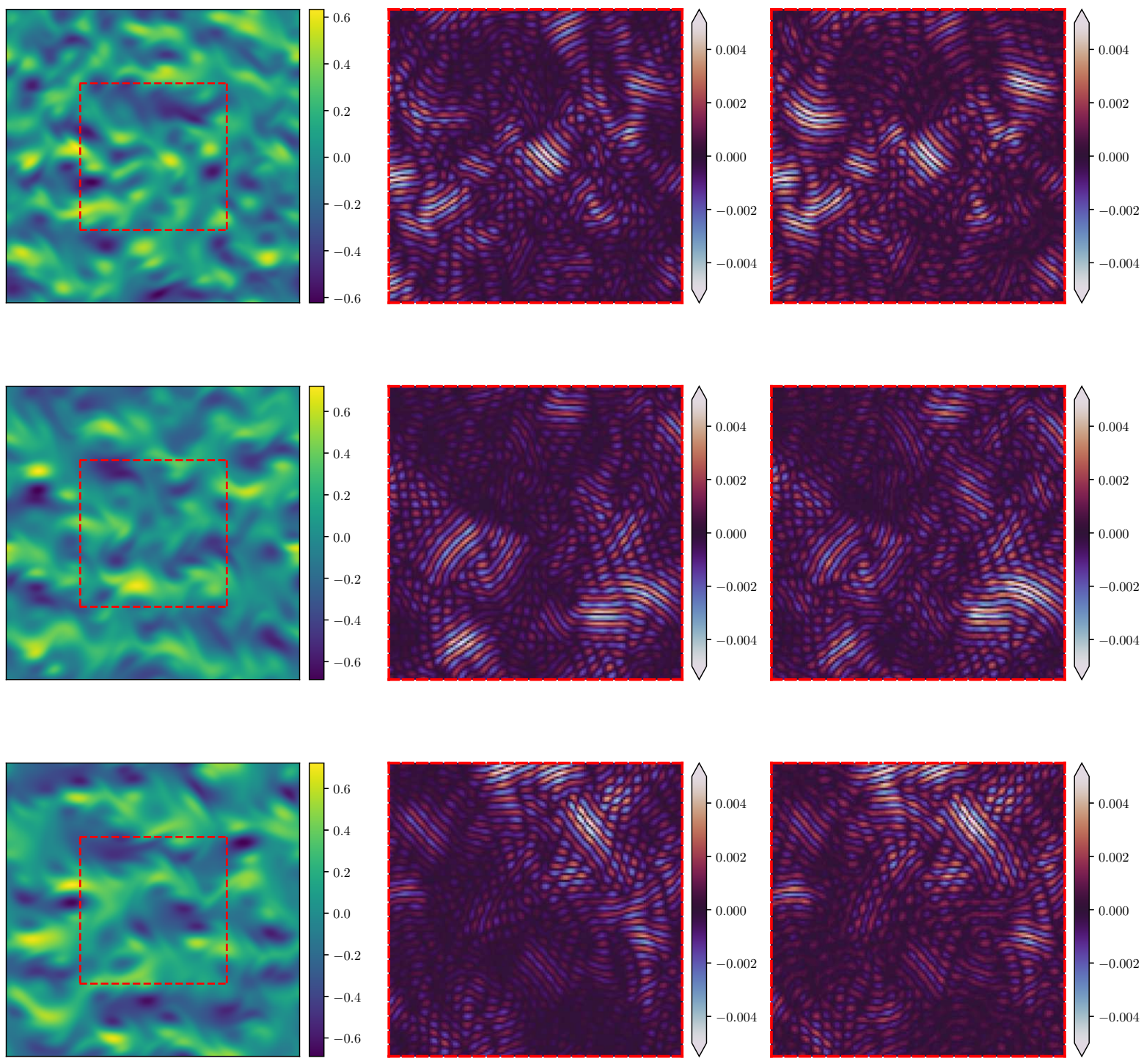

**FIG. 9:** $u_y\,(k \leq 40)_{\mathrm{GEN}}$ [left], $u_y\,(k > 40)_{\mathrm{DNS}}$ [middle], $u_y\,(k > 40)_{GEN}$ [right] at three representative timesteps. *GEN* is used for generated fields from the first and second stage diffusion models. Middle and right columns show fields that are centre-cropped for clarity – largest values occur nearer the boundaries and obscure main-field oscillations. The crop location is indicated on the low-wavenumber fields in the left column by the red dashed line, which corresponds to the boundary in the middle and right column.

The high-wavenumber flow fields are presented in Figure 9. The generated high-wavenumber fields show strong agreement with the ground truth DNS high-wavenumber fields. This lends credence to the conjecture that high-wavenumber recovery using diffusion models is made possible by isolating and rescaling high-wavenumber information. As before, we analyse spectral information to ascertain the turbulence recovery in wavenumber regions of interest, as in Figure 10.

The red dashed line indicated on Figure 10 shows the TKE spectrum for the first stage SR

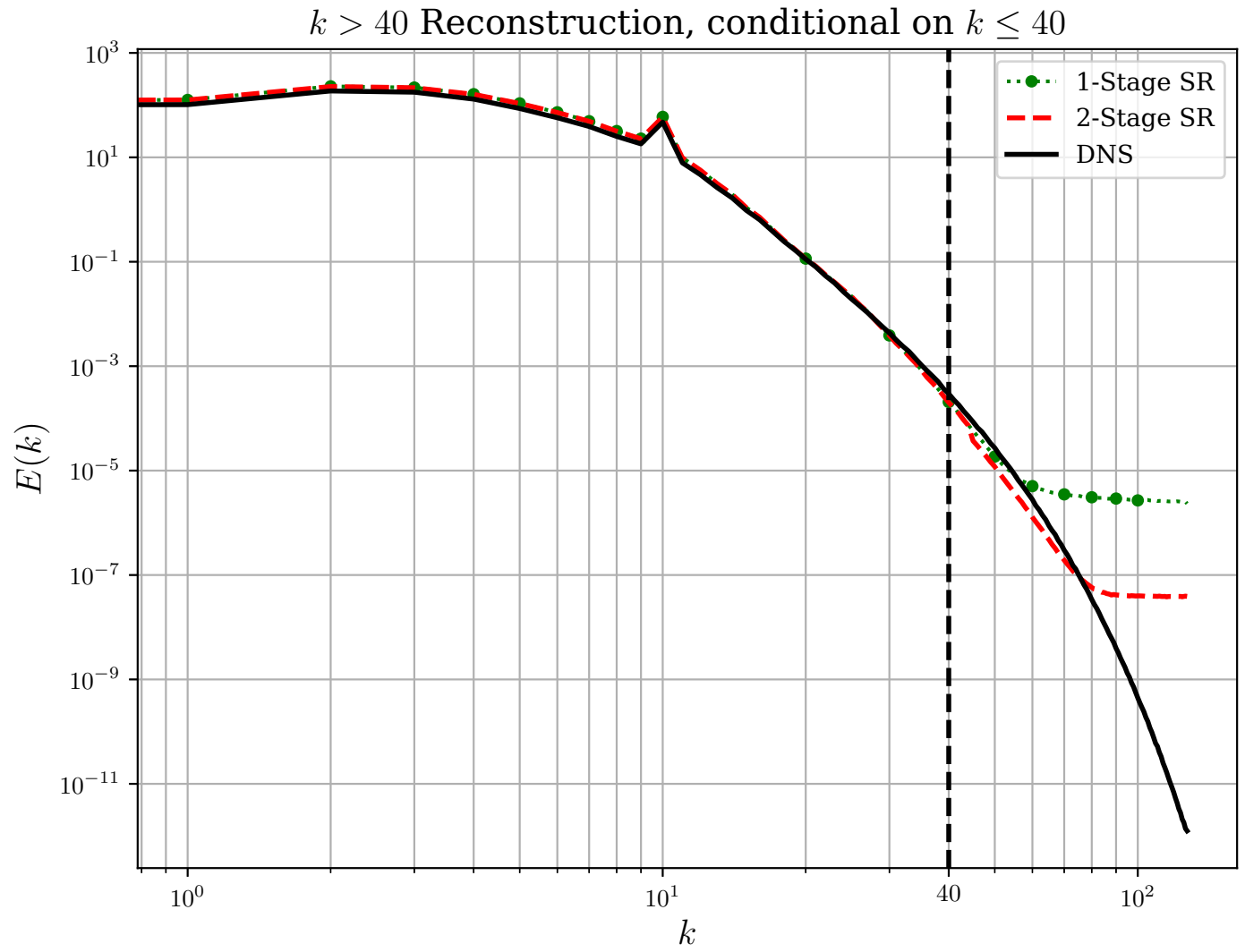

**FIG. 10:** TKE spectra from high-wavenumber conditional turbulence generation. The sum of low and high wavenumber components of the reconstructed flow is compared to the DNS field.

up to wavenumber $k = 40$, followed by the second stage generated high-wavenumber content for $k >= 40$. We demonstrate that our spectral decomposition diffusion model is able to recover turbulence information in this region up to $k = 80$, which means that through the first and second stage of turbulence generation, our method recovers a reasonable portion of the wavenumbers simulated in the DNS. We anticipate that a general N-stage model would iteratively be able to recover the whole spectrum, but leave this exercise as future work. The number of stages, N, may be considered a hyperparameter which in this study we fix to 2.

### C. Case 2: Spatial Super-Resolution of a Rayleigh-Bénard Convection Cell

To verify our approach on a more complex flow with body forces, we present results for a spatial super-resolution DDPM trained to learn the distribution of DNS temperature fields, $T$, given low-resolution representations of the fields. To our knowledge, this is the first piece of of work to carry out SR analysis on a Rayleigh-Bénard Convection cell at a high Rayleigh number of $Ra = 10^9$. $512 \times 1024$ fields are reconstructed from $64 \times 128$ samples of average-pooled DNS data. Instantaneous snapshots of scalar dissipation are discussed to qualitatively consider the turbulent mixing of the temperature field, PDFs are analysed to understand the distribution of predicted quantities, and spectra are analysed as before to investigate the length scales of turbulence recovered. Temperature fields considered here are in dimensionless units.

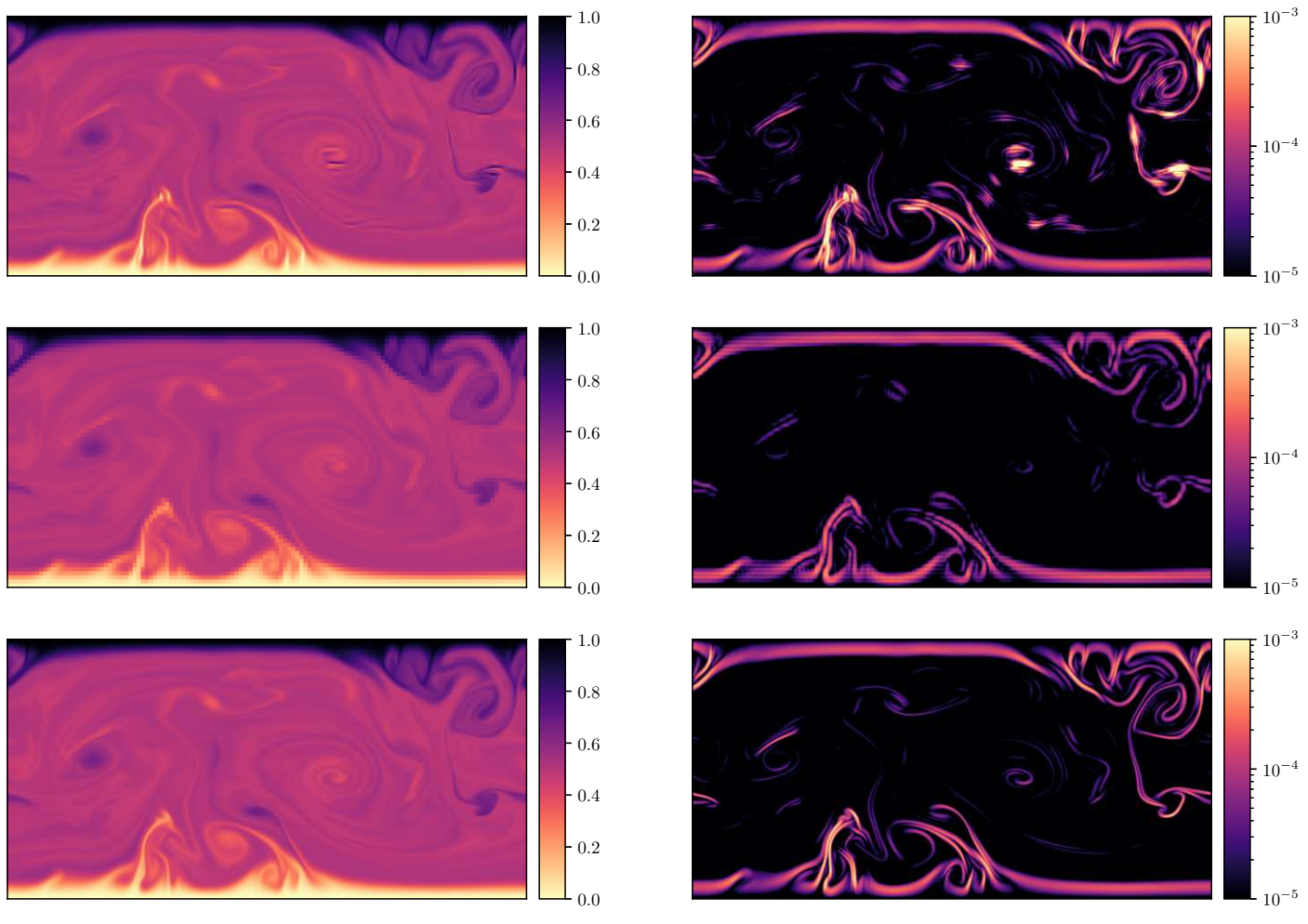

**FIG. 11:** Dimensionless temperature field (left), and corresponding scalar dissipation (right): (top) $512 \times 1024$ super-resolved, (middle) $64 \times 128$ low-resolution, (bottom) $512 \times 1024$ DNS.

The scalar dissipation is computed using a 2nd order central differencing scheme for the gradients, and is represented using a log normalised colormap in Figure 11. Recovery of high-frequency structures is observed in the super-resolved fields, with near-wall regions particularly well recovered. Structures which are only partially visible in the low-resolution are recovered in the super-resolved field (see right of samples). Side-by-side snapshots of the samples used to compute scalar dissipation show that the main field is recovered well. We note that the SR fields contain some spurious oscillations, which we believe to be an artefact of lower-capacity training. With increasing resolution, the parameterizing UNet should be made proportionally larger, which due to computational constraints, we were unable to do here.

We analyse the distribution of normalised temperature fluctuations, $T^*$, in Figure 12. The LR here is still relatively high-resolution, at $64 \times 128$, and hence the PDF matches the DNS well. We are encouraged to observe the super-resolved data also showing good agreement with the high-resolution data. This demonstrates the super-resolved fields are statistically similar to the ground truth DNS. We note that the SR-derived fluctuations have a very low probability of predicting values outside of the range of the input data, which is to be expected with generative models. Crucially, values around the mean correlate well with the DNS data, implying that the bulk temperature fluctuations are captured well by the DDPM output.

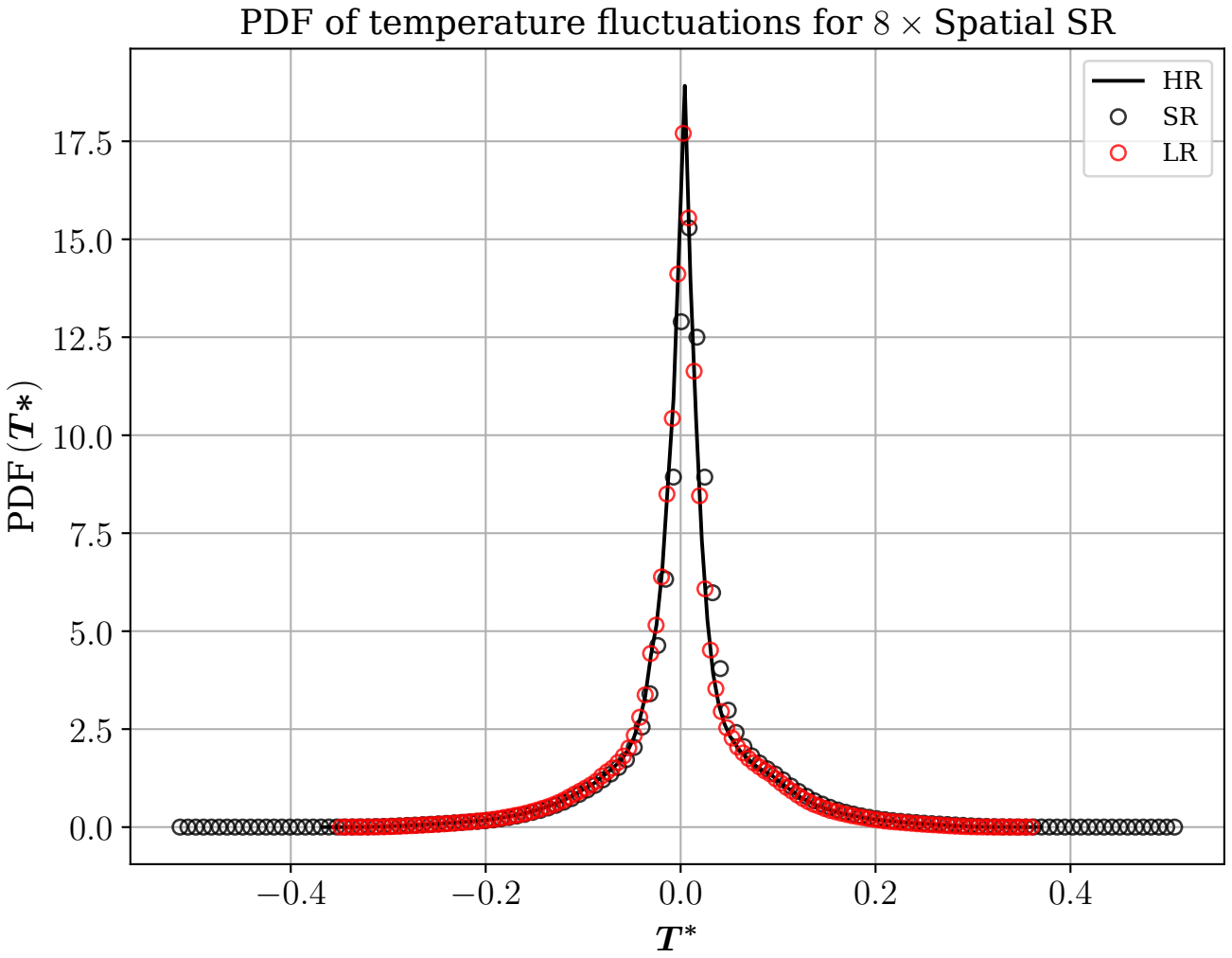

**FIG. 12:** PDF$(T^*)$ for the first stage, spatial super-resolution. LR is $64 \times 128$, HR is the $512 \times 1024$ DNS, and SR is from the super-resolved fields.

As the DNS resolution is large, it is difficult to discern missing high-frequency features in the downsampled data. Power spectral density plots provide a more rigorous means of examining the recovered fields, averaged over the validation set (Figure 13).

It is observed from Figure 13 that the DDPM-generated fields are statistically congruent with the DNS solution up to wavenumber $k = 60$, recovering a range of wavenumbers not present in the low-resolution field.

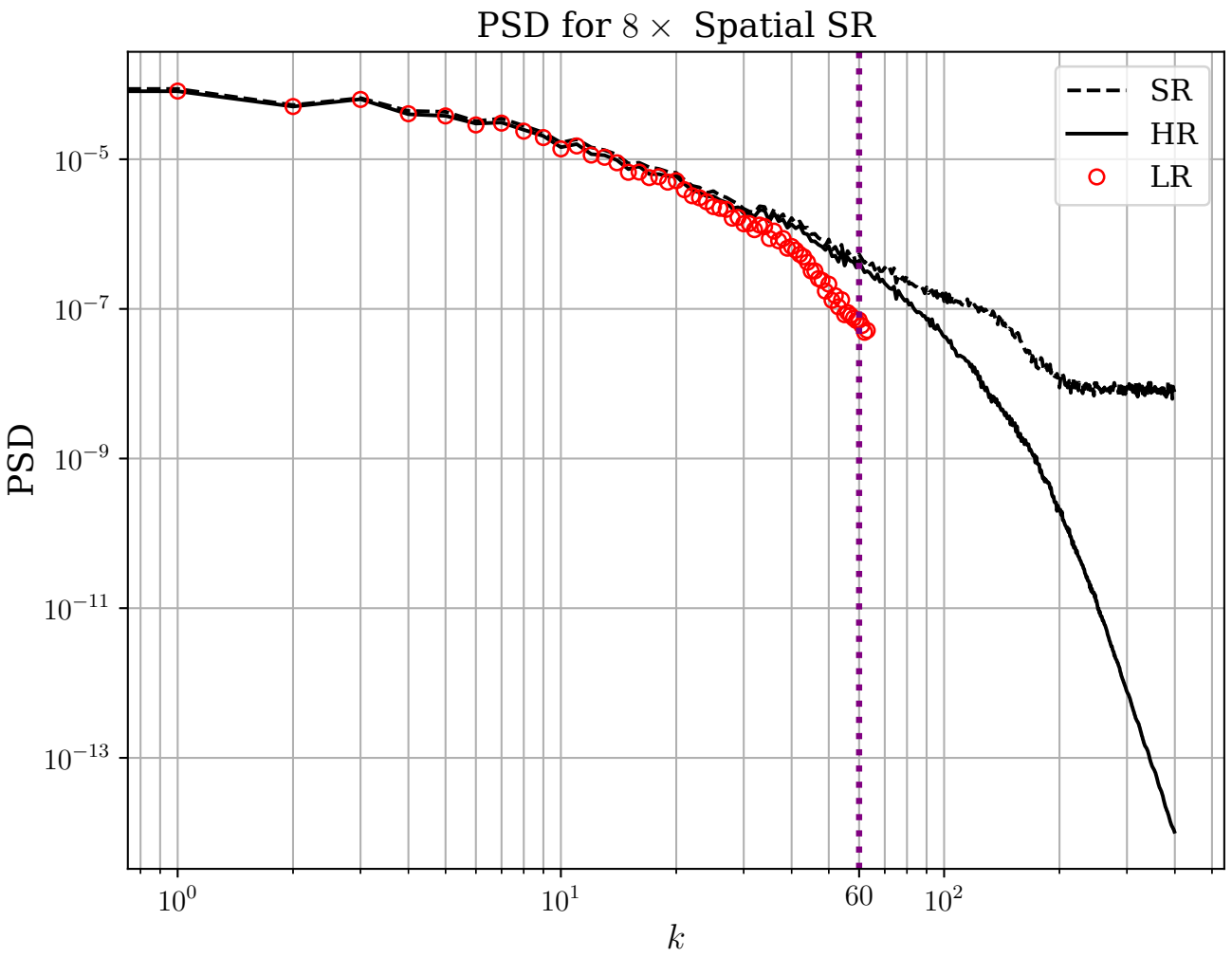

**FIG. 13:** Power Spectral Density for the $512 \times 1024$ DNS, $64 \times 128$ low-resolution, and $512 \times 1024$ DDPM-generated super-resolved fields.

### D. Case 2: Fourier-Filtered Turbulence Generation

As in III B, it is observed in the stage 1 results for the RBC case (III C), there is a limit up to which the DDPM is able to recover information. We present our findings on a similar high-wavenumber turbulence reconstruction task, filtering at $k = 60$.

Figure 14 highlights an instance of high-wavenumber turbulence recovery in the near-wall region. It is clearly observed that the stage 2 approach recovers the bulk structure of the high-wavenumber content. There is some missing detail in the DDPM-generated snapshot; we attribute this to the fact that the Rayleigh-Bénard Convection contains more complex physics than the Kolmogorov Flow, and the DNS is at a higher resolution. This implies that a greater portion of the spectrum needs recovery, through additional recursive stages of turbulence generation.

Generated fields are not produced deterministically. The purpose of using generative methods in turbulence is to recover statistics, rather than instantaneous fields. To this end, we analyse spectra in Figure 15, and note that our 2-stage turbulence recovery has significantly improved the range of wavenumbers which may be generated accurately (note the log-scale of the plot; the recovered wavenumbers improve from $k \simeq 60$ to $k \simeq 105$ – a 75% improvement). Again, we anticipate

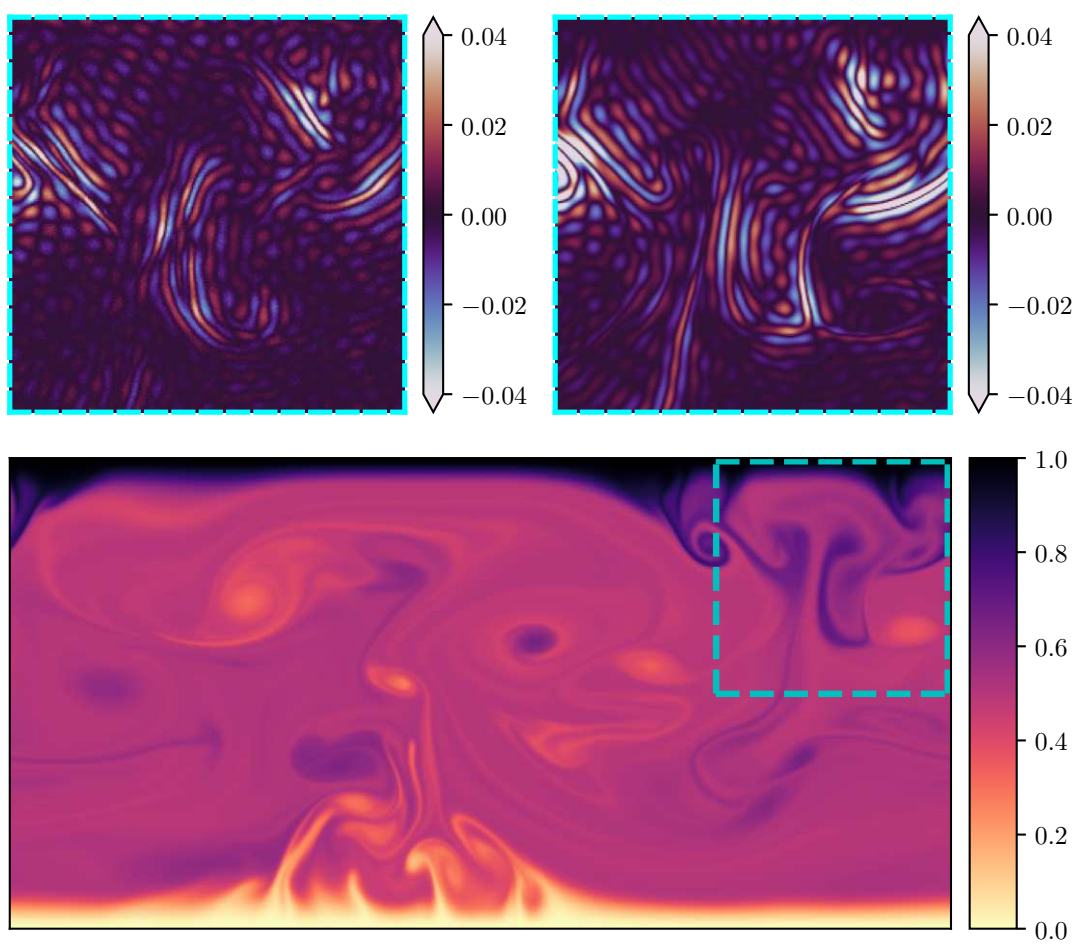

**FIG. 14:** An instantaneous snapshot of: (top left) DDPM-generated high-wavenumber content and (top right) DNS high-wavenumber content of a turbulent temperature field. The low-pass filtered field is included in the bottom figure.

arbitrary recovery via an iterative N-stage scheme but leave this as future work.

## IV. CONCLUSIONS, CONSIDERATIONS, AND FUTURE WORK

We have demonstrated that a diffusion model may be used as a generative technique to recover high-fidelity information lost from a high-resolution field. We extend approaches to do so by using spectral filters in wavenumber space to decompose flow fields into high and low wavenumber components, training a diffusion model to learn the conditional probability of a high-frequency field given a low-frequency field. We investigate the performance of our method by considering physical quantities recovered in the super-resolved flow. We find that diffusion models are a powerful generative method for super-resolution of turbulence, and that strong statistical recovery of turbulence can be achieved by the developed method.

We have omitted a discussion on inference cost in diffusion models, but a key open issue in this area is their iterative nature necessitates multiple network evaluations. This contrasts with other generative methods which require (typically) a single network evaluation through their parameterising network. Denoising Diffusion Implicit Models (DDIM) aim to address this by driving down the number of network evaluations through a novel sampling approach[39]. Additionally, we note that our two-stage turbulence recovery approach is equally applicable to other generative methods, such as GANs.

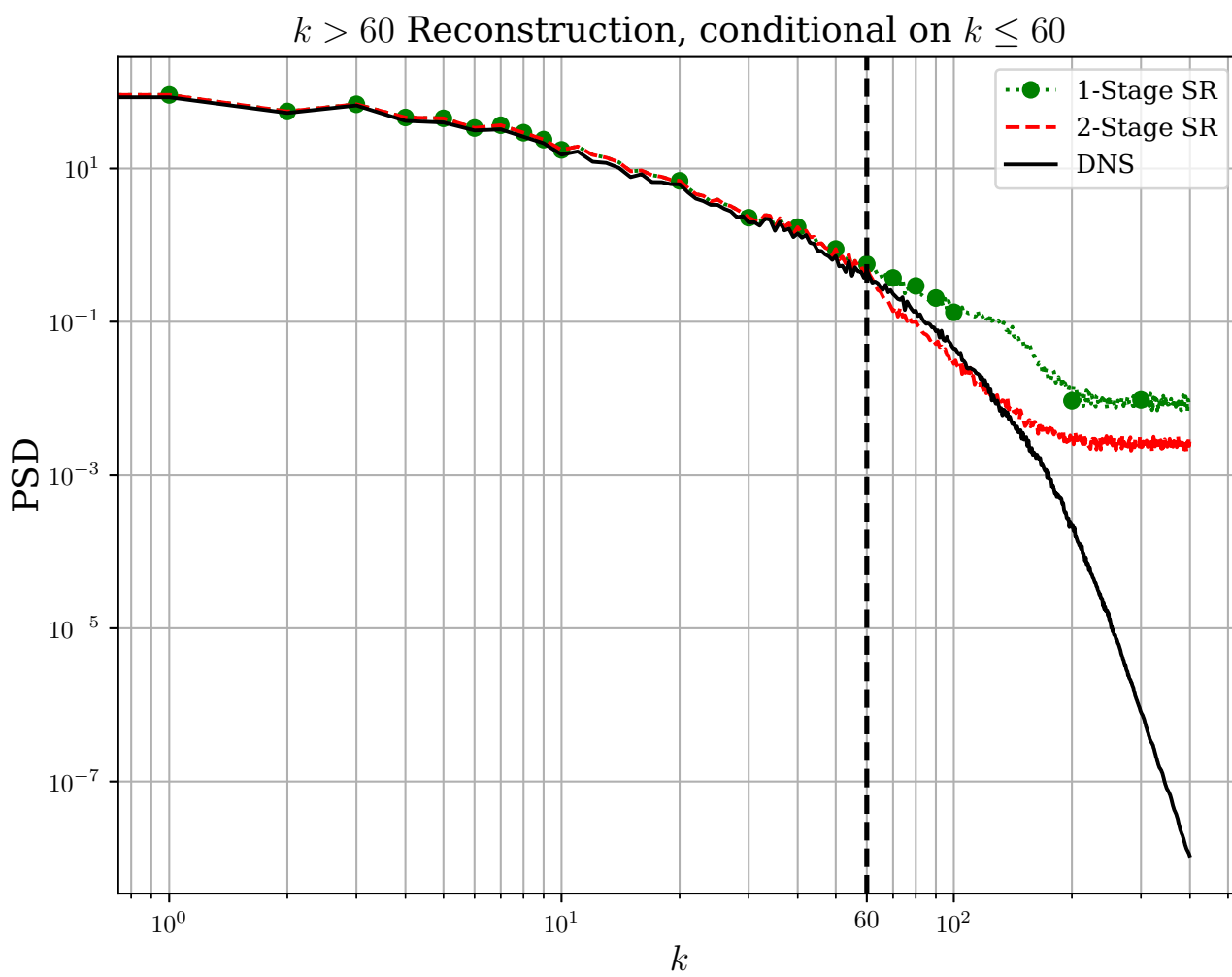

**FIG. 15:** Power Spectral Density for the DNS solution, the first stage spatial super-resolution DDPM, and the second stage spectrally decomposed DDPM.

A limitation of this approach to filtering in wavenumber space for a second stage of turbulence recovery is that the filter is dependent on the wavenumbers recovered in the first stage super-resolution. This has a direct impact on training time; the first- and second-stage diffusion models cannot be trained in parallel and must be trained sequentially after testing the performance of the first-stage spatial super-resolution model. We propose that our method may be extended by using a reverse process designed to recover specific wavenumbers, which may be known *a priori* and thus allowing for the second stage high-wavenumber diffusion model to be trained in parallel. This would also enable the use of variable wavenumber cutoffs for regions where the flow is resolved well vs regions where it is less well resolved.

Alternatively, from our experiments in spectral filtering, it was observed that progressively higher-wavenumber cutoffs lead to fields which are transformations of lower-wavenumber cutoffs, provided that the wavenumber filter is larger than the integral wavenumber. High-pass filtered fields are oscillations following the dominant energy carrying modes of a flow, and transformations between different amplitudes and frequencies of these oscillations reduce to a deterministic mapping problem, which may be simple to learn using a non-generative UNet. This would enable arbitrary high wavenumber turbulence generation, using our two-stage model to initially recover a large portion of the full DNS field, followed by a filter-transforming UNet to generate corresponding higher-wavenumber content.

We propose that this method could be extended to investigate specified ranges of wavenumber space. Spherical shells or doughnut shaped filters in 3D and 2D respectively allow for the decomposition of a flow into its constituent wavenumber modes. We believe that recursively applying corrections at different wavenumber ranges may be a viable option for subgrid-scale modelling using lower-fidelity CFD methods, where the maximum simulated wavenumber is less than the maximum wavenumber for a DNS-like solution.

## ACKNOWLEDGMENTS

This work was supported by the Engineering and Physical Sciences Research Council [EP-T517823-1]. The authors would like to acknowledge the assistance given by Research IT and the use of the Computational Shared Facility at The University of Manchester. The authors would also like to personally thank Daniel Kelshaw at Imperial College London for fruitful conversations regarding spectral decomposition. SD acknowledges a Dame Kathleen Ollerenshaw Fellowship.

## V. AUTHOR DECLARATION

The authors have no conflicts of interest to disclose.

## VI. DATA AVAILABILITY STATEMENT

The data that support the findings of this study are available from the corresponding author upon request. Supporting code can be found at `https://github.com/HamzaSardar/classifier-free-guidance/releases/tag/Submission`.

## REFERENCES

[1] Kai Fukami, Koji Fukagata, and Kunihiko Taira. Super-resolution reconstruction of turbulent flows with machine learning. *Journal of Fluid Mechanics*, 870:106–120, July 2019. ISSN 0022-1120, 1469-7645. doi:10.1017/jfm.2019.238.

[2] Bo Liu, Jiupeng Tang, Haibo Huang, and Xi-Yun Lu. Deep learning methods for super-resolution reconstruction of turbulent flows. *Physics of Fluids*, 32(2):025105, February 2020. ISSN 1070-6631. doi:10.1063/1.5140772.

[3] Zhideng Zhou, Binglin Li, Xiaolei Yang, and Zixuan Yang. A robust super-resolution reconstruction model of turbulent flow data based on deep learning. *Computers & Fluids*, 239:105382, May 2022. ISSN 00457930. doi:10.1016/j.compfluid.2022.105382.

[4] Tomoki Asaka, Katsunori Yoshimatsu, and Kai Schneider. Machine learning-based vorticity evolution and super-resolution of homogeneous isotropic turbulence using wavelet projection. *Physics of Fluids*, 36(2):025120, February 2024. ISSN 1070-6631, 1089-7666. doi: 10.1063/5.0185165. URL `https://pubs.aip.org/pof/article/36/2/025120/3262840/Machine-learning-based-vorticity-evolution-and`.

[5] Filippos Sofos, Dimitris Drikakis, Ioannis William Kokkinakis, and S. Michael Spottswood. A deep learning super-resolution model for turbulent image upscaling and its application to shock wave–boundary layer interaction. *Physics of Fluids*, 36(2):025117, February 2024. ISSN 1070-6631, 1089-7666. doi:10.1063/5.0190272. URL `https://pubs.aip.org/pof/article/36/2/025117/3262653/A-deep-learning-super-resolution-model-for`.

[6] Hyojin Kim, Junhyuk Kim, Sungjin Won, and Changhoon Lee. Unsupervised deep learning for super-resolution reconstruction of turbulence. *Journal of Fluid Mechanics*, 910:A29, March 2021. ISSN 0022-1120, 1469-7645. doi:10.1017/jfm.2020.1028.

[7] Claudia Drygala, Benjamin Winhart, Francesca di Mare, and Hanno Gottschalk. Generative Modeling of Turbulence. *Physics of Fluids*, 34(3):035114, March 2022. ISSN 1070-6631, 1089-7666. doi:10.1063/5.0082562.

[8] B. Galanti and A. Tsinober. Is turbulence ergodic? *Physics Letters A*, 330(3-4):173–180, September 2004. ISSN 03759601. doi:10.1016/j.physleta.2004.07.009.

[9] Zhiwen Deng, Chuangxin He, Yingzheng Liu, and Kyung Chun Kim. Super-resolution reconstruction of turbulent velocity fields using a generative adversarial network-based artificial intelligence framework. *Physics of Fluids*, 31(12):125111, December 2019. ISSN 1070-6631, 1089-7666. doi:

10.1063/1.5127031. URL https://pubs.aip.org/pof/article/31/12/125111/1075848/Super-resolution-reconstruction-of-turbulent.

[10]Mathis Bode, Michael Gauding, Zeyu Lian, Dominik Denker, Marco Davidovic, Konstantin Kleinheinz, Jenia Jitsev, and Heinz Pitsch. Using physics-informed enhanced super-resolution generative adversarial networks for subfilter modeling in turbulent reactive flows. *Proceedings of the Combustion Institute*, 38(2):2617–2625, 2021. ISSN 15407489. doi:10.1016/j.proci.2020.06.022. URL https://linkinghub.elsevier.com/retrieve/pii/S1540748920300481.

[11]Akshay Subramaniam, Man Long Wong, Raunak D. Borker, Sravya Nimmagadda, and Sanjiva K. Lele. Turbulence Enrichment using Physics-informed Generative Adversarial Networks, March 2020. URL http://arxiv.org/abs/2003.01907. arXiv:2003.01907 [physics].

[12]Mustafa Z. Yousif, Linqi Yu, and Hee-Chang Lim. Super-resolution reconstruction of turbulent flow fields at various Reynolds numbers based on generative adversarial networks. *Physics of Fluids*, 34(1):015130, January 2022. ISSN 1070-6631, 1089-7666. doi:10.1063/5.0074724. URL https://pubs.aip.org/pof/article/34/1/015130/2846421/Super-resolution-reconstruction-of-turbulent-flow.

[13]Qin Xu, Zijian Zhuang, Yongcai Pan, and Binghai Wen. Super-resolution reconstruction of turbulent flows with a transformer-based deep learning framework. *Physics of Fluids*, 35(5):055130, May 2023. ISSN 1070-6631, 1089-7666. doi:10.1063/5.0149551. URL https://pubs.aip.org/pof/article/35/5/055130/2890201/Super-resolution-reconstruction-of-turbulent-flows.

[14]Arvind T. Mohan, Dima Tretiak, Misha Chertkov, and Daniel Livescu. Spatio-temporal deep learning models of 3D turbulence with physics informed diagnostics. *Journal of Turbulence*, 21(9-10):484–524, October 2020. ISSN 1468-5248. doi:10.1080/14685248.2020.1832230. URL https://www.tandfonline.com/doi/full/10.1080/14685248.2020.1832230.

[15]Jonathan Ho, Ajay Jain, and Pieter Abbeel. Denoising Diffusion Probabilistic Models, December 2020.

[16]Zhisheng Xiao, Karsten Kreis, and Arash Vahdat. Tackling the Generative Learning Trilemma with Denoising Diffusion GANs, April 2022.

[17]Prafulla Dhariwal and Alex Nichol. Diffusion Models Beat GANs on Image Synthesis, June 2021.

[18]Dule Shu, Zijie Li, and Amir Barati Farimani. A physics-informed diffusion model for high-fidelity flow field reconstruction. *Journal of Computational Physics*, 478:111972, April 2023.

ISSN 00219991. doi:10.1016/j.jcp.2023.111972.

[19]Longzhang Huang, Chenxu Zheng, Yanyu Chen, Wenjiang Xu, and Fan Yang. Three-dimensional high-sampling super-resolution reconstruction of swirling flame based on physically consistent diffusion models. *Physics of Fluids*, 36(9):095113, September 2024. ISSN 1070-6631, 1089-7666. doi:10.1063/5.0225657. URL https://pubs.aip.org/pof/article/36/9/095113/3311569/Three-dimensional-high-sampling-super-resolution.

[20]Jonathan Ho and Tim Salimans. Classifier-Free Diffusion Guidance, July 2022. URL http://arxiv.org/abs/2207.12598. arXiv:2207.12598 [cs].

[21]Tianyi Li, Alessandra S. Lanotte, Michele Buzzicotti, Fabio Bonaccorso, and Luca Biferale. Multi-Scale Reconstruction of Turbulent Rotating Flows with Generative Diffusion Models. *Atmosphere*, 15(1):60, December 2023. ISSN 2073-4433. doi:10.3390/atmos15010060. URL https://www.mdpi.com/2073-4433/15/1/60.

[22]Yuzhuo Yin, Yuang Jiang, Mei Lin, and Qiuwang Wang. Velocity field reconstruction of mixing flow in T-junctions based on particle image database using deep generative models. *Physics of Fluids*, 36(8):085175, August 2024. ISSN 1070-6631, 1089-7666. doi:10.1063/5.0215252. URL https://pubs.aip.org/pof/article/36/8/085175/3309298/Velocity-field-reconstruction-of-mixing-flow-in-T.

[23]Tianyi Li, Luca Biferale, Fabio Bonaccorso, Martino Andrea Scarpolini, and Michele Buzzicotti. Synthetic Lagrangian Turbulence by Generative Diffusion Models. *Nature Machine Intelligence*, 6(4):393–403, April 2024. ISSN 2522-5839. doi:10.1038/s42256-024-00810-0. URL http://arxiv.org/abs/2307.08529. arXiv:2307.08529 [cond-mat, physics:nlin, physics:physics].

[24]Tianyi Li, Samuele Tommasi, Michele Buzzicotti, Fabio Bonaccorso, and Luca Biferale. Generative diffusion models for synthetic trajectories of heavy and light particles in turbulence. *International Journal of Multiphase Flow*, 181:104980, December 2024. ISSN 03019322. doi:10.1016/j.ijmultiphaseflow.2024.104980. URL https://linkinghub.elsevier.com/retrieve/pii/S030193222400257X.

[25]Georg Kohl, Li-Wei Chen, and Nils Thuerey. Benchmarking Autoregressive Conditional Diffusion Models for Turbulent Flow Simulation, January 2024. URL http://arxiv.org/abs/2309.01745. arXiv:2309.01745 [physics].

[26]Marten Lienen, David Lüdke, Jan Hansen-Palmus, and Stephan Günnemann. From Zero to Turbulence: Generative Modeling for 3D Flow Simulation, March 2024. URL http://arxiv.org/abs/2306.01776. arXiv:2306.01776 [physics].

[27] N. Platt, L. Sirovich, and N. Fitzmaurice. An investigation of chaotic Kolmogorov flows. *Physics of Fluids A: Fluid Dynamics*, 3(4):681–696, April 1991. ISSN 0899-8213. doi:10.1063/1.858074. URL `https://pubs.aip.org/pof/article/3/4/681/402184/An-investigation-of-chaotic-Kolmogorov-flows`.

[28] Chiyu lMaxr Jiang, Soheil Esmaeilzadeh, Kamyar Azizzadenesheli, Karthik Kashinath, Mustafa Mustafa, Hamdi A. Tchelepi, Philip Marcus, Mr Prabhat, and Anima Anandkumar. MESH-FREEFLOWNET: A Physics-Constrained Deep Continuous Space-Time Super-Resolution Framework. In *SC20: International Conference for High Performance Computing, Networking, Storage and Analysis*, pages 1–15, Atlanta, GA, USA, November 2020. IEEE. ISBN 978-1-72819-998-6. doi:10.1109/SC41405.2020.00013. URL `https://ieeexplore.ieee.org/document/9355293/`.

[29] Xinjie Wang, Siyuan Zhu, Yundong Guo, Peng Han, Yucheng Wang, Zhiqiang Wei, and Xiaogang Jin. TransFlowNet: A physics-constrained Transformer framework for spatio-temporal super-resolution of flow simulations. *Journal of Computational Science*, 65:101906, November 2022. ISSN 18777503. doi:10.1016/j.jocs.2022.101906. URL `https://linkinghub.elsevier.com/retrieve/pii/S1877750322002654`.

[30] Pu Ren, Chengping Rao, Yang Liu, Zihan Ma, Qi Wang, Jian-Xun Wang, and Hao Sun. PhySR: Physics-informed deep super-resolution for spatiotemporal data. *Journal of Computational Physics*, 492:112438, November 2023. ISSN 00219991. doi:10.1016/j.jcp.2023.112438. URL `https://linkinghub.elsevier.com/retrieve/pii/S0021999123005338`.

[31] Diane M. Salim, Blakesley Burkhart, and David Sondak. Extending a Physics-informed Machine-learning Network for Superresolution Studies of Rayleigh–Bénard Convection. *The Astrophysical Journal*, 964(1):2, March 2024. ISSN 0004-637X, 1538-4357. doi:10.3847/1538-4357/ad1c55. URL `https://iopscience.iop.org/article/10.3847/1538-4357/ad1c55`.

[32] Gideon Dresdner, Dmitrii Kochkov, Peter Norgaard, Leonardo Zepeda-Núñez, Jamie A. Smith, Michael P. Brenner, and Stephan Hoyer. Learning to correct spectral methods for simulating turbulent flows. 2022. doi:10.48550/ARXIV.2207.00556. URL `https://arxiv.org/abs/2207.00556`.

[33] Keaton J. Burns, Geoffrey M. Vasil, Jeffrey S. Oishi, Daniel Lecoanet, and Benjamin P. Brown. Dedalus: A flexible framework for numerical simulations with spectral methods. *Physical Review Research*, 2(2):023068, April 2020. ISSN 2643-1564. doi:10.1103/PhysRevResearch.2.023068. URL `https://link.aps.org/doi/10.1103/PhysRevResearch.2.023068`.

[34] Yann LeCun, Bernhard Boser, John Denker, Donnie Henderson, R. Howard, Wayne Hubbard, and Lawrence Jackel. Handwritten digit recognition with a back-propagation network. In D. Touretzky, editor, *Advances in Neural Information Processing Systems*, volume 2. Morgan-Kaufmann, 1989.

[35] Yann LeCun, Leon Bottou, Yoshua Bengio, and Patrick Ha. Gradient-Based Learning Applied to Document Recognition. 1998.

[36] Olaf Ronneberger, Philipp Fischer, and Thomas Brox. U-Net: Convolutional Networks for Biomedical Image Segmentation, May 2015.

[37] Daniel Kelshaw, Georgios Rigas, and Luca Magri. Physics-Informed CNNs for Super-Resolution of Sparse Observations on Dynamical Systems, November 2022.

[38] Sander Dieleman. Diffusion is spectral autoregression, September 2024. URL `https://sander.ai/2024/09/02/spectral-autoregression.html`.

[39] Jiaming Song, Chenlin Meng, and Stefano Ermon. Denoising Diffusion Implicit Models, October 2022.